\documentclass[sigconf]{acmart}

\usepackage{booktabs}
\usepackage{multirow}
\usepackage[normalem]{ulem}
\useunder{\uline}{\ul}{}
\usepackage[table,xcdraw]{xcolor}

\usepackage{graphicx}
\usepackage{subcaption}     

\usepackage{amsmath}
\usepackage{amsfonts}
\usepackage{amssymb}

\usepackage{balance}
\usepackage{amsthm}
\usepackage{enumitem}
\usepackage{color}

\usepackage{multirow}

\usepackage{bbding}

\usepackage{hyperref}
\usepackage{verbatim}

\AtBeginDocument{%
  }

\setcopyright{acmlicensed}
\copyrightyear{2018}
\acmYear{2018}
\acmDOI{XXXXXXX.XXXXXXX}
\acmConference[Conference acronym 'XX]{Make sure to enter the correct
  conference title from your rights confirmation email}{June 03--05,
  2018}{Woodstock, NY}
\acmISBN{978-1-4503-XXXX-X/2018/06}

\newcommand{\ie}{\emph{i.e., }}
\newcommand{\eg}{\emph{e.g., }}

\newcommand{\cf}{\emph{cf. }}
\newcommand{\aka}{\emph{a.k.a. }}

\begin{document}


\title{HatLLM: Hierarchical Attention Masking for Enhanced Collaborative Modeling in LLM-based Recommendation}



\author{Yu Cui}
\orcid{0009-0001-6203-3022}
\affiliation{%
  \institution{Zhejiang University}
  \city{Hangzhou}
  \country{China}}
\email{cuiyu23@zju.edu.cn}

\author{Feng Liu}
\orcid{0009-0004-9265-9431}
\affiliation{%
  \institution{OPPO Research Institute}
  \city{Shenzhen}
  \country{China}}
\email{liufeng4hit@gmail.com}

\author{Jiawei Chen}
\orcid{0000-0002-4752-2629}
\affiliation{%
  \institution{Zhejiang University}
  \city{Hangzhou}
  \country{China}}
\email{sleepyhunt@zju.edu.cn}
\authornote{Corresponding author.}

\author{Canghong Jin}
\affiliation{%
  \institution{Hangzhou City University}
  \city{Hangzhou}
  \country{China}}
\email{jinch@hzcu.edu.cn}

\author{Xingyu Lou}
\affiliation{%
  \institution{OPPO Research Institute}
  \city{Shenzhen}
  \country{China}}
\email{louxingyu@oppo.com}

\author{Changwang Zhang}
\affiliation{%
  \institution{OPPO Research Institute}
  \city{Shenzhen}
  \country{China}}
\email{changwangzhang@foxmail.com}

\author{Jun Wang}
\affiliation{%
  \institution{OPPO Research Institute}
  \city{Shenzhen}
  \country{China}}
\email{junwang.lu@gmail.com}

\author{Yuegang Sun}
\affiliation{%
  \institution{Intelligence Indeed}
  \city{Hangzhou}
  \country{China}}
\email{bulutuo@i-i.ai}

\author{Can Wang}
\affiliation{%
  \institution{Zhejiang University}
  \city{Hangzhou}
  \country{China}}
\email{wcan@zju.edu.cn}

\renewcommand{\shortauthors}{Cui et al.}

\begin{abstract}
Recent years have witnessed a surge of research on leveraging large language models (LLMs) for sequential recommendation. LLMs have demonstrated remarkable potential in inferring users' nuanced preferences through fine-grained semantic reasoning. However, they also exhibit a notable limitation in effectively modeling collaborative signals, \ie, behavioral correlations inherent in users' historical interactions. Our empirical analysis further reveals that the attention mechanisms in LLMs tend to disproportionately focus on tokens within the same item, thereby impeding the capture of cross-item correlations.

To address this limitation, we propose a novel hierarchical attention masking strategy for LLM-based recommendation, termed HatLLM. Specifically, in shallow layers, HatLLM masks attention between tokens from different items, facilitating intra-item semantic understanding; in contrast, in deep layers, HatLLM masks attention within items, thereby compelling the model to capture cross-item correlations. This progressive, layer-wise approach enables LLMs to jointly model both token-level and item-level dependencies. Extensive experiments on three real-world datasets demonstrate that HatLLM achieves significant performance gains (9.13\% on average) over existing LLM-based methods.

\end{abstract}

\ccsdesc[500]{Information systems~Recommender systems}

\keywords{Sequential Recommendation, Large language Model}


\maketitle

\section{Introduction}

\begin{figure*}[t]
    \centering
    \includegraphics[width=\textwidth]{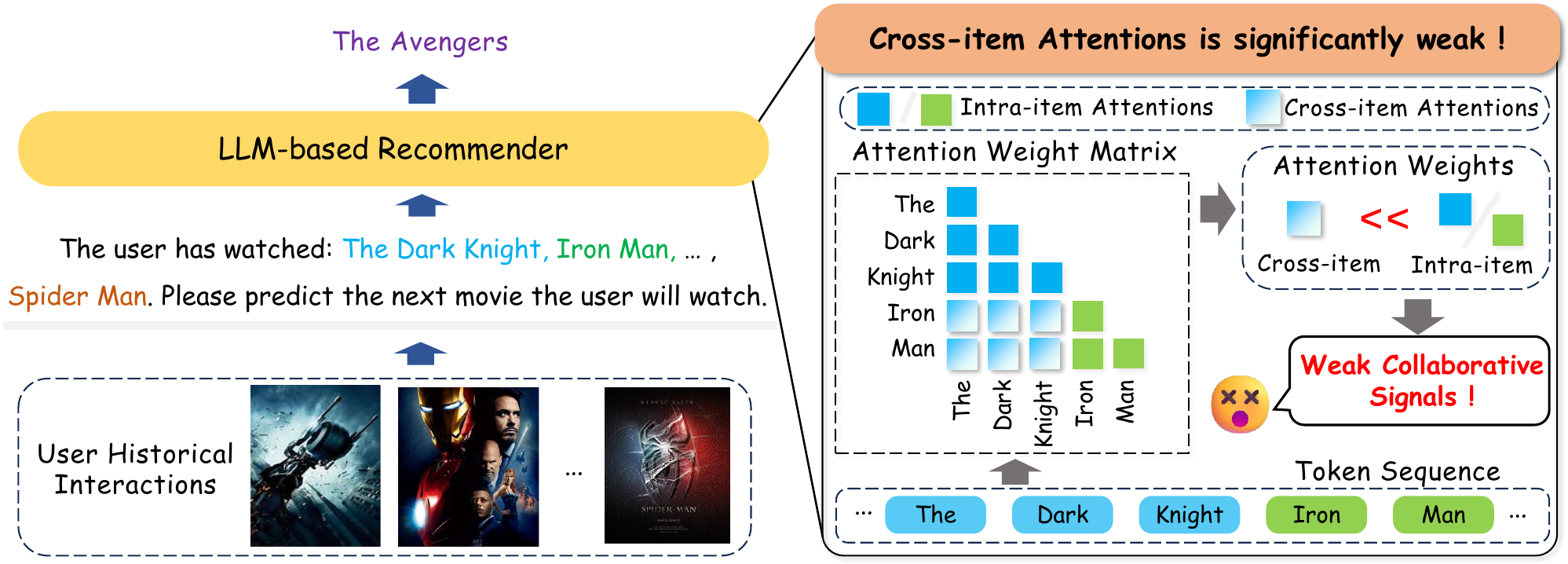}
    \caption{Illustration of LLM-based recommendation methods and their limitations in capturing cross-item correlations.} 
    \label{fig:intro}
\end{figure*}


Sequential Recommendation (SR) aims to predict users' next items of interest by analyzing their historical interaction sequences~\cite{kang2018self,xie2022contrastive,yang2023generic}. Traditional SR models typically employ sequential neural networks (\eg Transformers~\cite{khan2022transformers} or LSTMs~\cite{yu2019review}), capturing behavioral correlations (\aka collaborative information) inherent in users' historical interactions to deduce their evolving preferences. 

The advent of large language models (LLMs), with their remarkable open-domain knowledge and semantic reasoning capabilities, has spurred interest in adapting LLMs for sequential recommendation~\cite{bao2023tallrec,hua2023index,zheng2024adapting,lin2024bridging,rajput2023recommender,wang2025msl}. These approaches often reformulate SR as a language modeling task, where items are translated into textual descriptions (\eg titles) and organized into prompts to guide LLMs in predicting future interactions. By reasoning at a fine-grained semantic token level, LLMs offer the potential to capture subtle semantic patterns in user interests and thus represent a promising direction for advancing recommender systems.

Despite their strengths, LLM-based recommendation faces a fundamental limitation: \textbf{ineffective modeling of collaborative signals}~\cite{liao2024llara,zhang2025collm,kim2024large}. Our empirical analysis reveals that token-level processing inherently limits LLMs' ability to capture cross-item correlations. Specifically, when analyzing attention weights across user histories, we find that the aggregated attention between tokens from different items is rather weak, exhibiting values significant smaller than the attentions within the same item (\eg about three times lower on Amazon datasets). This phenomenon is prevalent and likely stems from the attention skew effect: the attention mechanism of LLMs exhibits a bias toward intra-item tokens due to their closer positional relationships and higher co-occurrence frequency in training datasets, thereby impeding the modeling of cross-item correlations. This observation motivates a key research question: \textbf{\textit{How can we enable LLMs to effectively model collaborative signals for sequential recommendation?}}


Existing attempts address this by explicitly injecting item embeddings learned from traditional models as special tokens to supplement collaborative signals~\cite{liao2024llara,zhang2025collm,kim2024large}. However, these approaches face two key limitations: (1) The rich collaborative information inherent in user histories is compressed into low-dimensional embeddings, inevitably incurring substantial information loss; (2) The influence of such tokens is often diluted by the large volume of surrounding tokens, resulting in limited performance gains.

To overcome these challenges, we propose a novel \underline{\textbf{H}}ierarchical \underline{\textbf{at}}tention masking strategy for \underline{\textbf{LLM}}-based Recommendation, named \textbf{HatLLM}. HatLLM adopts distinct masking strategies across LLM layers to model diverse information: 1) In shallow layers, HatLLM masks attention between tokens of different items, facilitating the understanding of intra-item semantics; 2) In middle layers, HatLLM retains the original attention mechanism, preserving fine-grained semantic modeling across all tokens; 3) In deep layers, HatLLM masks intra-item attention, compelling the model to focus on holistic cross-item correlations. This progressive, layer-wise approach --- from fine-grained semantic understanding to coarse-grained item-level modeling --- empowers LLMs to jointly capture token-level and item-level dependencies for improved recommendation.

Notably, HatLLM is easily implemented, requiring only minimal modifications to the LLM’s attention masking scheme without introducing additional overhead. Extensive experiments on three benchmark datasets demonstrate that HatLLM surpasses state-of-the-art LLM-based recommenders, achieving an average improvement of 9.13\%.

In summery, our contributions are as follows:
\begin{itemize}[leftmargin=*]
\item We provide empirical analyses that reveal the limitations of existing LLMs in modeling collaborative information for recommendation.
\item We introduce HatLLM, a novel hierarchical attention masking strategy that enables LLMs to model both token-level and item-level correlations at different layers.
\item We conduct comprehensive experiments to validate the effectiveness of HatLLM over existing LLM-based recommendation approaches.
\end{itemize}

\begin{figure}[t]
    \centering
    \begin{subfigure}{0.23\textwidth}
        \centering
        \includegraphics[width=\textwidth]{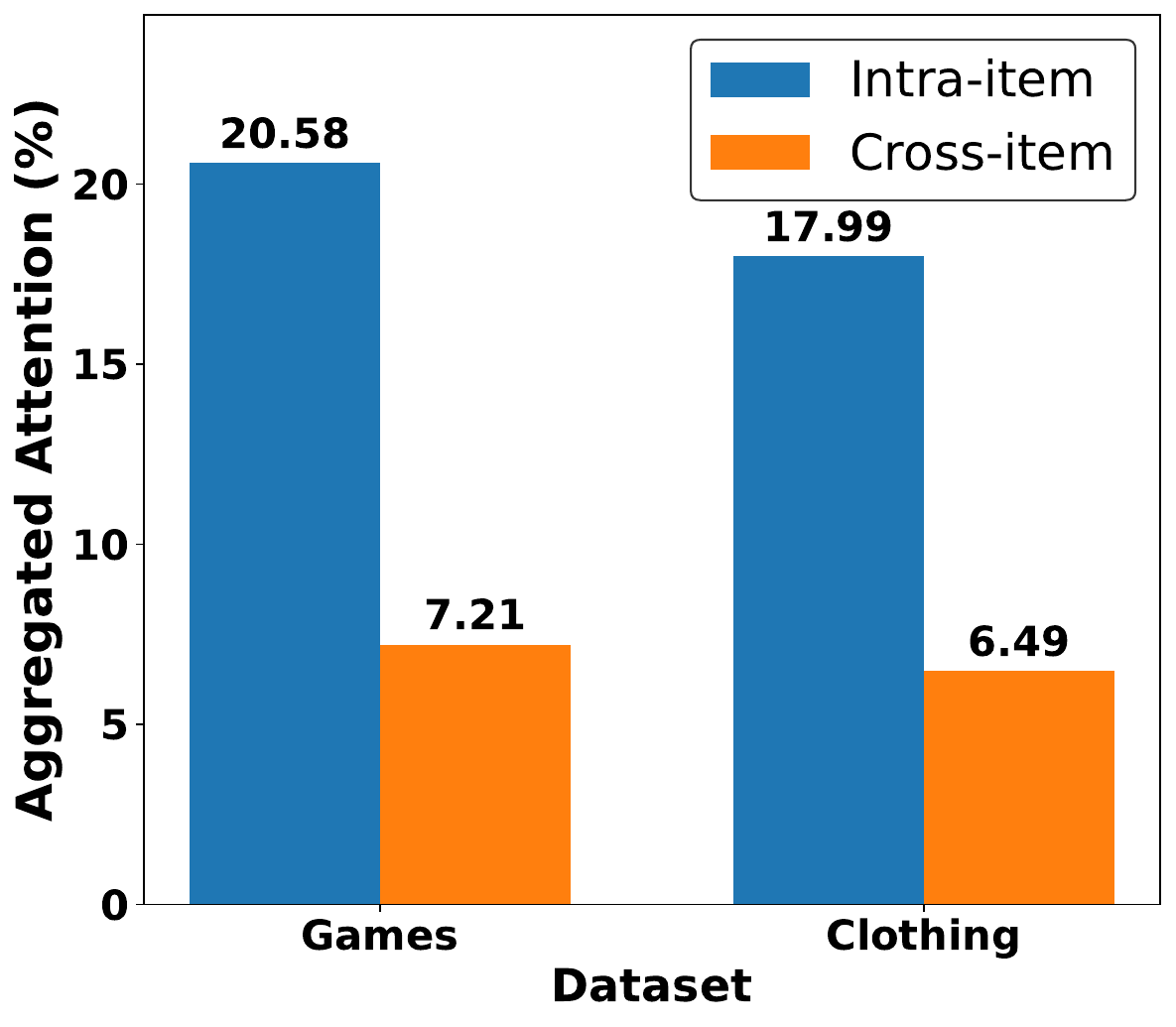}
    \end{subfigure}    
    \begin{subfigure}{0.23\textwidth}
        \centering
        \includegraphics[width=\textwidth]{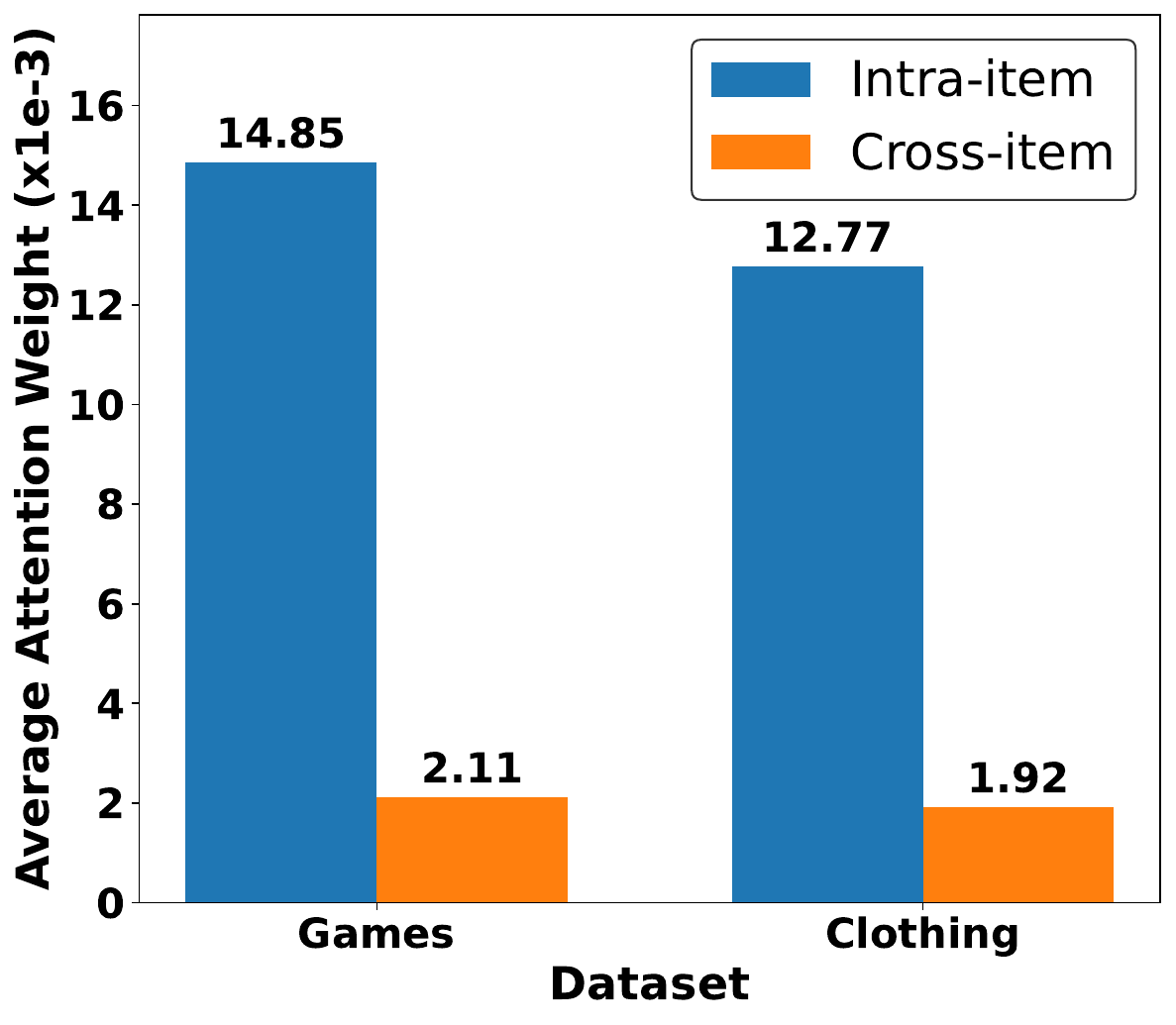}
    \end{subfigure}
    \vspace{-0.3cm}
    \caption{The total attention proportion and average attention weight comparison on intra-item and inter-item attentions.} 
    \label{fig:imbalence}
    \vspace{-0.3cm}
\end{figure}

\section{Preliminary}
\subsection{Task Formulation} 
This work studies on sequential recommendation~\cite{boka2024survey}, which has secured a pivotal role in various modern recommendation systems. Sequential recommendation aims to deduce user dynamic preferences based on their history interactions. With a user set $\mathcal{U}$ and an item set $\mathcal{I}$,  a user's historical interactions can be organized in chronological order $ S_u = (i_1, i_2, \ldots, i_{t-1})$ where $i_{k} \in \mathcal{I}$ represents the $k$-th item that the user $u$ interacted with. The task of sequential recommendation is to predict the next item $i_{t}$ that the user is likely to interact with. 

\subsection{LLM-based recommendation} 
Given the sophisticated semantic reasoning capabilities of Large Language Models, LLM-based recommendation systems~\cite{lin2024rella,xi2024play,zhang2025collm,chen2024hllm,lin2024clickprompt} have attracted considerable attention in sequential recommendation research. These approaches typically reformulate sequential recommendation (SR) as a language modeling task, comprising three primary components:

\textbf{1) Prompt Construction.} As illustrated in Figure~\ref{fig:intro}, LLM-based recommendation systems first represent items through their textual descriptions (\eg, titles), then organize users' historical interactions into structured prompts to guide LLM predictions.

\textbf{2) Item Prediction.} Recent LLM-based studies mainly employ two strategies for item retrieval: (i) \textit{language-based strategy}, which directly generates natural language descriptions of the next item for item grounding~\cite{bao2025bi}; and (ii) \textit{embedding-based strategy}, which leverages LLMs to generate item embeddings, subsequently mapped to the item space through a learnable projection head for next-item prediction~\cite{li2023e4srec,xu2025slmrec}.

While both strategies are prevalent in recent literature, considering the significant computational overhead of language generation, we simply adopt the embedding-based strategy for our proposed method. For baseline comparisons, we implement both strategies and report the superior performance for each method.

\textbf{3) Supervised Fine-tuning.} Since LLMs are not inherently optimized for recommendation tasks, supervised fine-tuning has proven essential for task alignment. This process pairs constructed prompts with target items for optimization. Corresponding to the two prediction strategies, distinct optimization objectives are employed: (i) for language-based prediction~\cite{bao2023tallrec,bao2025bi}, the language modeling loss is utilized, optimizing the generative likelihood of tokens in target item descriptions; (ii) for embedding-based prediction~\cite{li2023e4srec,xu2025slmrec}, cross-entropy loss is applied to align predicted item embeddings with ground-truth representations.

\subsection{Attention Mechanism in LLMs} 

Existing Large Language Models (LLMs) are mainly built upon the Transformer architecture~\cite{vaswani2017attention}.  The attention mechanism constitutes a fundamental component of LLM-based methods, enabling the capture of contextual dependencies between tokens. Formally, let $X=[x_1,x_2,..x_n]$ denote the sequence of token embeddings at a specific layer, where  $n$ represents the sequence length.  These token embeddings are then projected into three matrices: Query (\textbf{Q}), Key (\textbf{K}), and Value (\textbf{V}). Subsequently, each token computes a weighted aggregation over all preceding tokens through the following attention mechanism:

\begin{equation}
\textbf{O} = \text{Attention}(\textbf{Q},\textbf{K},\textbf{V}) \triangleq \text{softmax}\left( \frac{\textbf{Q}\textbf{K}^{\top}}{\sqrt{d}} + \textbf{M} \right) \textbf{V},
\end{equation}
\begin{equation}
M_{jk} \triangleq 
\begin{cases}
0, & \text{if}\quad j \geq k \\
-\infty, & \text{otherwise}
\end{cases}
\end{equation}

where \(d\) is the dimension of the key matrix used for scaling the dot products, $j$ and $k$ represent the $j$-th token and $k$-th token in the token sequence. The attention weights are determined by the scaled dot product between queries and keys, quantifying the pairwise relevance between tokens. The matrix $M$ serves as a causal masking matrix, ensuring each token only attends to preceding tokens in the sequence. This autoregressive property is essential for maintaining causality during generation~\cite{achiam2023gpt,dubey2024llama}. The attention mechanism effectively captures sequential dependencies between tokens, contributing the remarkable success of LLMs across various tasks.

\begin{figure*}[t]
    \centering 
    \includegraphics[width=\textwidth]{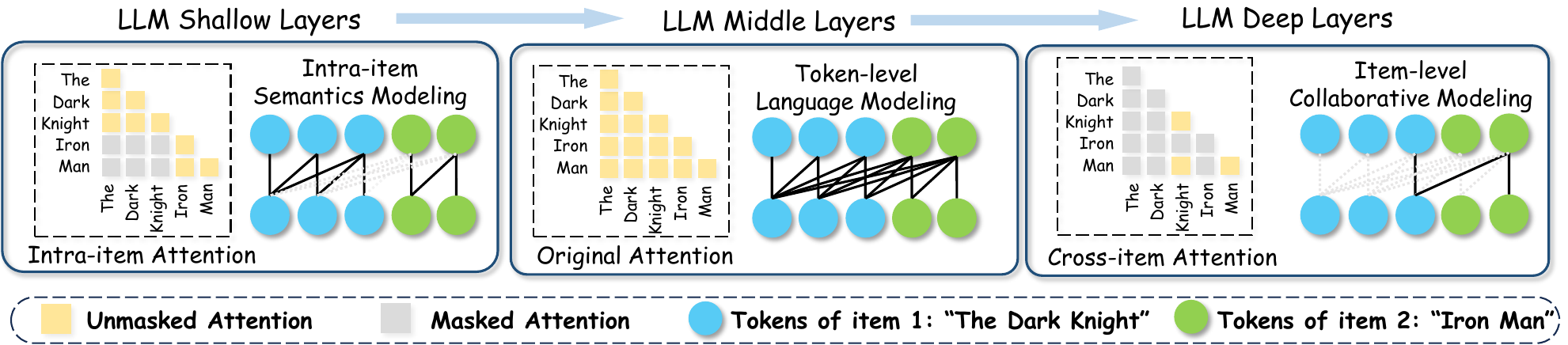}
    \vspace{-0.3cm}
    \caption{The overall framework of proposed HatLLM. It contains: 1) Intra-item Attention (IN) in shallow layers for learning individual item semantics; 2) Original Attention (OR) in middle layers for token-level language modeling; 3) Cross-item Attention (CR) in deep layers for capturing collaborative signals of cross-item correlations.} 
    \label{fig:framework} 
    \vspace{-0.3cm}
\end{figure*}

\begin{figure}[t]
    \centering 
    \includegraphics[width=0.48\textwidth]{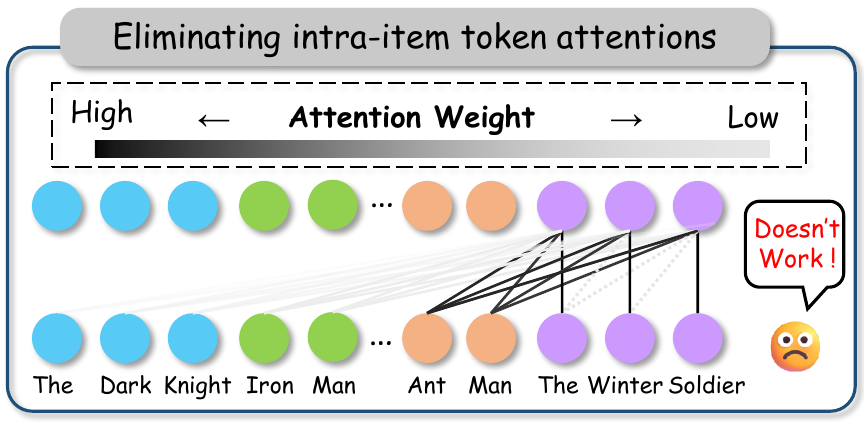}
    \vspace{-0.3cm}
    \caption{Illustration of the limitation of directly masking intra-item token attention with $M^{CR-pre}_{jk}$. Here we give an example of the token attention aggregation of the item "The Winter Soldier": attention easily focus on nearby items.} 
    \label{fig:IntraAM} 
    \vspace{-0.3cm}
\end{figure}

\begin{table}[]
\centering
\caption{The prompt template for instruction tuning.}
\vspace{-0.3cm}
\label{tab:prompt}
\begin{tabular}{@{}ll@{}}
\toprule
\multicolumn{2}{c}{Instruction   Input}                           \\ \midrule
\multicolumn{1}{l|}{Instruction:} & Given a list of movies the user has   watched before, \\
\multicolumn{1}{l|}{} & please recommend a new movie that the user \\
\multicolumn{1}{l|}{} & likes to the user.                        \\ \midrule
\multicolumn{1}{l|}{Input:}       & The user has watched the following   movies before:    \\
\multicolumn{1}{l|}{} & "The Dark Knight", "Iron Man", …          \\ 
\midrule
\multicolumn{2}{c}{Instruction   Output}                        \\ \midrule
\multicolumn{1}{l|}{Output:} & {"The Avengers"}        \\ 
\bottomrule
\end{tabular}
\vspace{-0.3cm}
\end{table}

\section{Analyses on Attention Weights}

While token-level processing of user historical interaction sequences is commonly employed in LLM-based recommendation methods~\cite{bao2023tallrec,bao2025bi,liao2024llara}, this approach inherently constrains LLMs' capacity to capture    cross-item correlations. To substantiate this claim, we conduct comprehensive analyses of attention weight distributions across tokens. Specifically, we fine-tuned BIGRec, a representative LLM-based recommendation method, on benchmark datasets including Amazon Games and Clothing. We subsequently computed the cumulative attention weights between tokens from different items and compared these values with intra-item token attention weights. Figure~\ref{fig:imbalence} presents the results, from which we draw the following key observation:

\textit{
\textbf{Observation:} The aggregated attention weights between tokens from different items are substantially weaker than those within the same item.
}

This finding demonstrates a significant limitation of LLMs in modeling collaborative information: LLMs tend to focus on intra-item correlations while struggling to capture cross-item dependencies. Furthermore, given that the number of token pairs spanning different items vastly exceeds that of intra-item pairs, the average attention weight value per cross-item token pair is much higher (\eg over seven times) than that of intra-item pairs. This undesirable phenomenon is pervasive and can be attributed to two primary factors: 

1) Positional Proximity Effect: Recent studies have shown that the attention mechanism in LLMs exhibits a strong bias toward tokens that are close in position~\cite{jiang2025beyond,hou2024large}. Since tokens within the same item are typically positioned closer together compared to tokens from different items --- for instance, for example, the average intra-item token distance is 5.6 and 4.8 for the Games and Clothing datasets, respectively, which is much smaller than the cross-item token distances (30.03 and 29.42) --- the attention is naturally concentrated on intra-item tokens.

2) Training Data Bias:  In sequential recommendation, LLMs are frequently fine-tuned on task-specific data. Notably, the co-occurrence frequency of intra-item token pairs in the training data is substantially higher than that of cross-item pairs (\eg 6.8 vs. 2.3 on Games and 8.1 vs. 2.5 on Clothing, respectively).
As a result, LLMs tend to capture stronger intra-item correlations during fine-tuning. This bias is further evidenced by the reduced cross-item attention weights in fine-tuned models relative to their pre-trained counterparts, as shown in Figure~\ref{fig:intro}.

Given these limitations, a critical question emerges: \textit{How can we enable LLMs to effectively model collaborative signals for sequential recommendation?} Several approaches have attempted to address this challenge by explicitly injecting collaborative item embeddings~\cite{lin2024rella,zhang2025collm}.  These methods typically pre-train traditional recommendation models to derive item embeddings, which are subsequently incorporated as special tokens in prompts to supplement collaborative information. However, we contend that this strategy suffers from two fundamental limitations: 1) These approaches rely on auxiliary traditional methods to extract and compress collaborative information into low-dimensional embeddings, inevitably incurring information loss. Besides, their effectiveness are subjected to the capabilities of the traditional methods. 2) given the substantial volume of other tokens, the influence of these collaborative tokens becomes severely diluted. To empirically validate this claim, we analyzed LLaRA~\cite{liao2024llara}, a representative method in this category. The average attention scores from collaborative tokens are remarkably low (1.85\% and 1.14\% on Games and Clothing datasets, respectively).  While LLaRA demonstrates performance improvements, the gains remain modest (3.80\% on average, \cf Table~\ref{tab:main}). Therefore, novel strategies that directly address the inherent attention bias of LLMs, rather than relying on external models, warrant investigation.

\section{Methodology}
In this section, we first present the proposed hierarchical attention masking strategy, HatLLM, and subsequently discuss its relationship to and distinctions from recent work.

\subsection{Hierarchical Attention Masking Strategy}

HatLLM applies distinct attention masking schemes across different layers of the LLM to facilitate the modeling of both fine-grained semantic information and coarse-grained item-level dependencies:

\textbf{Shallow Layers --- Intra-item Attention. } Comprehensive semantic understanding of individual items forms the foundation of effective recommendations. In the shallow layers, HatLLM masks attention between tokens belonging to different items, enabling the model to focus on modeling the intrinsic semantics of each item in the user interaction sequence without mutual interference, thereby enhancing their semantic understanding on items. 

Formally, this can be implemented by introducing an additional attention masking matrix  $M^{IN}_{jk}$ in equation (1) with:
\begin{equation}
M^{IN}_{jk} \triangleq 
\begin{cases}
-\infty, &  F(j)=1 \, \wedge  \, F(k)=1 \, \wedge  \, B(j,k)=0 \\ 
0, & \text{otherwise}
\end{cases}
\end{equation}
where $F(i)$ is an indicator denoting whether the token is an item token; and $B(j,k)$  denotes whether the $j$-th and $k$-th token belong to the same item. The introduction of $M^{Intra}_{jk}$ effectively masks cross-item attention.

\textbf{Middle Layers --- Original Attention. } In the middle layer, we preserve the original attention mechanism over all tokens. This design choice aims to maintain the original semantic modeling capability of LLM-based recommendations, capturing token-level correlations to infer users' nuanced semantic-level preferences.

\textbf{Deep Layers --- Cross-item Attention.} Given the inherent limitations of LLMs in capturing cross-item correlations, our analyses presented in Section 3 reveal that the major cause stems from excessive attention on intra-item tokens. To address this challenge, a natural approach is to directly mask the attention between intra-item tokens, enabling LLMs to focus on inferring correlations across items. Specifically, we introduce another distinct attention masking matrix $M^{CR-pre}_{jk}$ in equation (1) with:
\begin{equation}
M^{CR-pre}_{jk} \triangleq 
\begin{cases}
-\infty, &   j\ne k \, \wedge \, F(j)=1 \, \wedge \, F(k)=1 \, \wedge \, B(j,k)=1 \\ 
0, & \text{otherwise}
\end{cases}
\end{equation}

where intra-item token attention has been eliminated. However, in practice, we find this direct modification does not yield significant performance gains and may occasionally result in slightly degraded performance (\cf Table~\ref{tab:ab}, comparing "HatLLM" and "CR -> CR-pre"). This phenomenon can be attributed to the following: As illustrated in Figure~\ref{fig:IntraAM}, when we mask intra-item attention, the attention weights predominantly concentrate on tokens from nearby items. For example, the token 'Man' primarily aggregates information from tokens 'The', 'Dark', and 'Knight'. This pattern continues to impede LLMs from capturing relatively long-range item dependencies. Furthermore, given the potentially disparate semantics of adjacent items in the sequence, aggregating excessive information from neighboring items may adversely affect model performance.

To address this issue, a natural solution is to preserve only the last token from each item for cross-item attention modeling. This approach reduces the number of tokens available for aggregation, thereby facilitating the modeling of long-term item dependencies. Additionally, uniformly reducing each item's representation to a single token prevents attention bias toward items with more tokens. Furthermore, this design naturally emulates the item-wise dependency modeling of traditional methods --- the last token of an item has already aggregated the item's semantics and other relevant semantic information from preceding tokens, thus serving as a natural semantic representation of the item. Performing attention among these last tokens can therefore be interpreted as information propagation between items, effectively capturing collaborative signals. Formally, this straightforward approach can be readily implemented by modifying the previous mask matrix as follows:
\begin{equation}
\small
M^{CR}_{jk} \triangleq 
\begin{cases}
-\infty, &  j\ne k \, \wedge \, F(j)=1 \, \wedge \, F(k)=1 \, \wedge \, (L(j)=0\, \vee \,L(k)=0) \\ 
0, & \text{otherwise}
\end{cases}
\end{equation}
where we introduce a function $L(j)$ that judge whether the $j$-th token is the last token of an item. 

In summary, through this progressive, layer-wise architecture --- from fine-grained semantic understanding to coarse-grained item-level modeling --- LLMs can simultaneously capture token-level semantic information and item-level collaborative signals to achieve superior recommendation performance.

\begin{table}[t]
\centering
\caption{Statistics of the datasets.}
\vspace{-0.3cm}
\label{tab:datasets}
\begin{tabular}{@{}c|ccc@{}}
\toprule
\textbf{Dataset} & \textbf{Game} & \textbf{Beauty} & \textbf{Clothing} \\ \midrule
\#User & 54955 & 22332 & 39230 \\
\#Item & 17250 & 12086 & 22948 \\
\#Interaction & 416813 & 198215 & 277534 \\
Density & 0.0440\% & 0.0734\% & 0.0308\% \\ \bottomrule
\end{tabular}
\vspace{-0.3cm}
\end{table}

\subsection{Discussions}

\textbf{Comparison with Lite-LLM4Rec~\cite{wang2024rethinking}.}  Lite-LLM4Rec employs an item LLM to encode item semantics and generate item representations, followed by a user LLM that encodes these item representations to infer user preferences. This strategy appears similar to the combination of our intra-item attention layers and cross-item attention layers. However, we highlight two key distinctions between our method and theirs: 1) We implement intra-item and cross-item modeling within a single LLM using diverse attention mechanisms, enabling end-to-end fine-tuning of both components to better capture recommendation-specific dependencies. 2) We additionally preserve the middle layers with original attention among all semantic tokens, thereby maintaining LLMs' fine-grained semantic-level modeling of user preferences.

 \textbf{Comparison with Direct Stacking of LLMs and Traditional Models.} Another related strategy involves directly stacking LLMs with traditional models --- \ie replacing our cross-attention layers with traditional models. Specifically, one might first capture token-level patterns using LLMs and then extract semantic-enhanced item embeddings from LLMs to augment traditional recommendation models. Given the embedding dimension mismatch between LLMs and traditional methods, a learnable projector~\cite{su2012linear} or dimension reduction strategy~\cite{mackiewicz1993principal} can be employed. However, we emphasize that our strategy offers significant advantages: 1) HatLLM operates as an integrated system and can be optimized end-to-end, with all components fine-tuned jointly to maximize recommendation performance, whereas LLM and traditional models are difficult to optimize jointly. 2) Due to the different embedding spaces of LLMs and traditional models, bridging these two models inevitably requires embedding transformation and compression, which may result in information loss. Our ablation studies demonstrate that HatLLM significantly outperforms this stacking strategy (\cf Table~\ref{tab:ab}, comparing "HatLLM" with "CR->SASRec").

 \textbf{Comparing with DiscRec~\cite{liu2025discrec}.} DiscRec\footnote{DiscRec is excluded from our experiments as it targets a different problem setting beyond LLM-based recommendation. Our attempts to adapt it to our experiments yielded unsatisfactory results. } also aims to model both token-level semantic and item-level collaborative signals. However, HatLLM differs from DiscRec in two fundamental aspects: 1) Different scenarios: HatLLM is designed for LLM-based recommendation, while DiscRec targets generative recommendation; 2) Different architectures: HatLLM operates within one LLM by modifying the attention mechanism, while DiscRec employs parallel branches that fuse basic generative recommendation models with traditional collaborative models.

\section{Experiments}
We aim to answer the following research questions:
\begin{itemize}[leftmargin=*]
  \item $\mathbf{RQ1:}$ How does HatLLM perform compare to existing state-of-the-art recommendation methods?
  \item $\mathbf{RQ2:}$ What are the impacts of different components of HatLLM on its performance?
  \item $\mathbf{RQ3:}$ How do hyperparameters influence HatLLM?
\end{itemize}

\subsection{Experimental Settings}
\subsubsection{Datasets}
Three conventional real-world datasets: \textit{Amazon Video Games}, \textit{Amazon Beauty} and \textit{Amazon Clothing, Shoes and Jewelry} are utilized in our experiments~\footnote{\url{https://cseweb.ucsd.edu/~jmcauley/datasets/amazon/links.html}}, which are commonly used for the studies of LLM-based recommendation~\cite{bao2023tallrec,cui2024distillation}. 
For fair comparisons, we closely adhered to the preprocessing methods used in recent work~\cite{bao2025bi,wang2025msl,cui2024distillation}. Specifically, we firstly apply the 5-core setting to the original dataset, where users/items with fewer than 5 interactions will be filtered out. After that, a sliding window of length 11 is applied to segment the sequences. The resulting sequences are then sorted in ascending order by timestamp and split into training, validation, and testing sets with an 8:1:1 ratio. This approach ensures that the predicted interactions during testing come after all interactions observed during training, preventing the data leakage~\cite{bao2025bi,wang2025msl}. The dataset statistics are presented in Table~\ref{tab:datasets}.




\subsubsection{Baselines}
The following methods are compared: 
\begin{itemize}[leftmargin=*]
  \item \textbf{Traditional recommenders: SASRec (ICDM'18)~\cite{kang2018self}, DROS (SIGIR'23)~\cite{yang2023generic}.} SASRec utilizes a self-attention-based model to model the user interaction sequence. DROS incorporates Distributionally Robust Optimization (DRO) to improve the model’s resilience to distributional shifts.
  \item \textbf{LLM-enhanced recommenders: KAR (RecSys'24)~\cite{xi2024towards}, LLM-CF (CIKM'24)~\cite{sun2024large}, DLLM2Rec (RecSys'24)~\cite{cui2024distillation}.} KAR utilizes LLM's open-world knowledge to get enhanced semantic users/items embeddings. LLM-CF leverage LLMs to generate the Chain-of-Thought (CoT) knowledge, which are further retrieved to enhance traditional RS models. DLLM2Rec introduces a knowledge distillation method to enhance the performance of traditional recommenders.
  \item \textbf{LLM-based recommenders: E4SRec (arXiv'23)~\cite{li2023e4srec}, Lite-LLM4Rec (arXiv'24)~\cite{wang2024rethinking}, LLaRA (SIGIR'24)~\cite{liao2024llara}, BIGRec (TORS'25)~\cite{bao2025bi}, SLMRec (ICLR'25)~\cite{xu2025slmrec}.} These are well-known and representative LLM-based recommendation methods with different strategies. Readers may refer to the related work and the appendix for more details about these methods. 
\end{itemize}


\subsubsection{Evaluation Metrics}
We employed two widely-used metrics $HR@K$ and $NDCG@K$ to evaluate the performance ($K=5,10$) following previous work~\cite{wang2025msl}. HR (Hit Ratio) evaluates whether a relevant item appears in the top-N recommendations. NDCG (Normalized Discounted Cumulative Gain) evaluates recommendation quality accounting for both relevance scores and their positions.

\begin{table*}[]
\centering
\caption{The performance comparison on three real-world datasets. The best and the second performances are bolded and underlined, respectively. "Impr." denotes the improvement of HatLLM over the best baseline method. "N" represents NDCG, and "H" represents Hit Ratio.}
\vspace{-0.3cm}
\label{tab:main}
\scalebox{0.85}{
\begin{tabular}{@{}cc|cccc|cccc|cccc@{}}
\toprule
\multicolumn{2}{c|}{} & \multicolumn{4}{c|}{\textbf{Games}} & \multicolumn{4}{c|}{\textbf{Beauty}} & \multicolumn{4}{c}{\textbf{Clothing}} \\ \cmidrule(l){3-14} 
\multicolumn{2}{c|}{\multirow{-2}{*}{\textbf{Method}}} & \textbf{H@5} & \textbf{N@5} & \textbf{H@10} & \textbf{N@10} & \textbf{H@5} & \textbf{N@5} & \textbf{H@10} & \textbf{N@10} & \textbf{H@5} & \textbf{N@5} & \textbf{H@10} & \textbf{N@10} \\ \midrule
\multicolumn{1}{c|}{} & SASRec (ICDM'18) & 0.0272 & 0.0155 & 0.0386 & 0.0191 & 0.0152 & 0.0080 & 0.0242 & 0.0109 & 0.0074 & 0.0039 & 0.0126 & 0.0056 \\
\multicolumn{1}{c|}{\multirow{-2}{*}{\textbf{Traditional}}} & DROS (SIGIR'23) & 0.0290 & 0.0181 & 0.0418 & 0.0222 & 0.0182 & 0.0102 & 0.0300 & 0.0140 & 0.0088 & 0.0055 & 0.0138 & 0.0071 \\ \midrule
\multicolumn{1}{c|}{} & KAR (RecSys'24) & 0.0292 & 0.0193 & 0.0464 & 0.0229 & 0.0182 & 0.0104 & 0.0290 & 0.0139 & 0.0094 & 0.0065 & 0.0136 & 0.0075 \\
\multicolumn{1}{c|}{} & LLM-CF (CIKM'24) & 0.0298 & 0.0199 & 0.0468 & 0.0229 & 0.0183 & 0.0105 & 0.0301 & 0.0143 & 0.0096 & 0.0066 & 0.0134 & 0.0078 \\
\multicolumn{1}{c|}{\multirow{-3}{*}{\textbf{LLM-enhanced}}} & DLLM2Rec (RecSys'24) & 0.0322 & 0.0227 & 0.0484 & 0.0278 & 0.0194 & 0.0113 & 0.0338 & 0.0159 & 0.0098 & 0.0065 & 0.0144 & 0.0080 \\ \midrule
\multicolumn{1}{c|}{} & Lite-LLMRec   (arXiv'24) & 0.0242 & 0.0172 & 0.0362 & 0.0210 & 0.0174 & 0.0107 & 0.0280 & 0.0141 & 0.0100 & 0.0061 & 0.0180 & 0.0086 \\
\multicolumn{1}{c|}{} & E4SRec (arXiv'23) & 0.0284 & 0.0198 & 0.0384 & 0.0236 & 0.0174 & 0.0097 & 0.0280 & 0.0134 & 0.0084 & 0.0042 & 0.0140 & 0.0061 \\
\multicolumn{1}{c|}{} & SLMRec (ICLR'25) & 0.0290 & 0.0209 & 0.0408 & 0.0246 & 0.0180 & 0.0101 & 0.0266 & 0.0125 & 0.0078 & 0.0041 & 0.0134 & 0.0059 \\
\multicolumn{1}{c|}{} & BIGRec (TORS'25) & 0.0344 & 0.0246 & 0.0498 & 0.0296 & 0.0230 & 0.0152 & 0.0360 & 0.0193 & 0.0138 & 0.0087 & 0.0214 & 0.0111 \\
\multicolumn{1}{c|}{} & LLaRA (SIGIR'24) & {\ul 0.0349} & {\ul 0.0251} & {\ul 0.0502} & {\ul 0.0302} & {\ul 0.0246} & {\ul 0.0173} & {\ul 0.0368} & {\ul 0.0213} & {\ul 0.0140} & {\ul 0.0089} & {\ul 0.0216} & {\ul 0.0113} \\
\multicolumn{1}{c|}{} & \cellcolor[HTML]{96FFFB}HatLLM & \cellcolor[HTML]{96FFFB}\textbf{0.0374} & \cellcolor[HTML]{96FFFB}\textbf{0.0275} & \cellcolor[HTML]{96FFFB}\textbf{0.0514} & \cellcolor[HTML]{96FFFB}\textbf{0.0320} & \cellcolor[HTML]{96FFFB}\textbf{0.0280} & \cellcolor[HTML]{96FFFB}\textbf{0.0194} & \cellcolor[HTML]{96FFFB}\textbf{0.0406} & \cellcolor[HTML]{96FFFB}\textbf{0.0235} & \cellcolor[HTML]{96FFFB}\textbf{0.0156} & \cellcolor[HTML]{96FFFB}\textbf{0.0091} & \cellcolor[HTML]{96FFFB}\textbf{0.0252} & \cellcolor[HTML]{96FFFB}\textbf{0.0122} \\ \cmidrule(l){2-14} 
\multicolumn{1}{c|}{\multirow{-7}{*}{\textbf{LLM-based}}} & \cellcolor[HTML]{96FFFB}Impr. & \cellcolor[HTML]{96FFFB}7.16\% & \cellcolor[HTML]{96FFFB}9.46\% & \cellcolor[HTML]{96FFFB}2.39\% & \cellcolor[HTML]{96FFFB}5.83\% & \cellcolor[HTML]{96FFFB}13.82\% & \cellcolor[HTML]{96FFFB}12.24\% & \cellcolor[HTML]{96FFFB}10.33\% & \cellcolor[HTML]{96FFFB}10.54\% & \cellcolor[HTML]{96FFFB}11.43\% & \cellcolor[HTML]{96FFFB}1.91\% & \cellcolor[HTML]{96FFFB}16.67\% & \cellcolor[HTML]{96FFFB}7.81\% \\ \bottomrule
\end{tabular}
}

\end{table*}

\begin{table}[ht]
\centering
\caption{The performance on LLMs of other parameter scales.}
\vspace{-0.3cm}
\label{tab:8b}
\scalebox{0.88}{
\begin{tabular}{@{}cc|cc|cc@{}}
\toprule
\multicolumn{2}{c|}{\multirow{2}{*}{\textbf{Method}}} & \multicolumn{2}{c|}{\textbf{Games}} & \multicolumn{2}{c}{\textbf{Clothing}} \\ \cmidrule(l){3-6} 
\multicolumn{2}{c|}{} & \textbf{H@10} & \textbf{N@10} & \textbf{H@10} & \textbf{N@10} \\ \midrule
\multicolumn{1}{c|}{\multirow{5}{*}{\textbf{Llama3-1B}}} & E4SRec (TORS'25) & 0.0372 & 0.0233 & 0.0124 & 0.0052 \\
\multicolumn{1}{c|}{} & SLMRec (ICLR'25) & 0.0382 & 0.0237 & 0.0112 & 0.0053 \\
\multicolumn{1}{c|}{} & BIGRec (TORS'25) & 0.0436 & 0.0259 & 0.0181 & 0.0098 \\
\multicolumn{1}{c|}{} & LLaRA   (SIGIR'24) & 0.0438 & 0.0260 & 0.0183 & 0.0099 \\
\multicolumn{1}{c|}{} & HatLLM & \textbf{0.0452} & \textbf{0.0263} & \textbf{0.0186} & \textbf{0.0101} \\ \midrule
\multicolumn{1}{c|}{\multirow{5}{*}{\textbf{Llama3-3B}}} & E4SRec (TORS'25) & 0.0384 & 0.0236 & 0.0140 & 0.0061 \\
\multicolumn{1}{c|}{} & SLMRec (ICLR'25) & 0.0408 & 0.0246 & 0.0134 & 0.0059 \\
\multicolumn{1}{c|}{} & BIGRec (TORS'25) & 0.0498 & 0.0296 & 0.0214 & 0.0111 \\
\multicolumn{1}{c|}{} & LLaRA   (SIGIR'24) & 0.0502 & 0.0302 & 0.0216 & 0.0113 \\
\multicolumn{1}{c|}{} & HatLLM & \textbf{0.0514} & \textbf{0.0320} & \textbf{0.0252} & \textbf{0.0122} \\ \midrule
\multicolumn{1}{c|}{\multirow{5}{*}{\textbf{Llama3-8B}}} & E4SRec (TORS'25) & 0.0386 & 0.0238 & 0.0148 & 0.0066 \\
\multicolumn{1}{c|}{} & SLMRec   (ICLR'25) & 0.0409 & 0.0248 & 0.0138 & 0.0065 \\
\multicolumn{1}{c|}{} & BIGRec   (TORS'25) & 0.0494 & 0.0304 & 0.0248 & 0.0129 \\
\multicolumn{1}{c|}{} & LLaRA (SIGIR'24) & 0.0498 & 0.0305 & 0.0256 & 0.0131 \\
\multicolumn{1}{c|}{} & HatLLM & \textbf{0.0534} & \textbf{0.0324} & \textbf{0.0268} & \textbf{0.0139} \\ \bottomrule
\end{tabular}
\vspace{-0.3cm}
}
\end{table}

\begin{table}[]
\centering
\caption{Ablation study on HatLLM. }
\vspace{-0.3cm}
\label{tab:ab}
\scalebox{0.9}{
\begin{tabular}{@{}c|cc|cc|cc@{}}
\toprule
\multirow{2}{*}{\textbf{Method}} & \multicolumn{2}{c|}{\textbf{Games}} & \multicolumn{2}{c|}{\textbf{Clothing}} & \multicolumn{2}{c}{\textbf{Beauty}} \\ \cmidrule(l){2-7} 
 & \textbf{H@10} & \textbf{N@10} & \textbf{H@10} & \textbf{N@10} & \textbf{H@10} & \textbf{N@10} \\ \midrule
BIGRec & 0.0498 & 0.0296 & 0.0214 & 0.0111 & 0.0360 & 0.0193 \\
HatLLM & \textbf{0.0514} & \textbf{0.0320} & \textbf{0.0252} & \textbf{0.0122} & \textbf{0.0406} & \textbf{0.0235} \\ \midrule
w/o IN & 0.0511 & 0.0317 & 0.0222 & 0.0112 & 0.0352 & 0.0203 \\
w/o OR & 0.0502 & 0.0302 & 0.0224 & 0.0115 & 0.0352 & 0.0209 \\
w/o CR & 0.0510 & 0.0311 & 0.0224 & 0.0112 &  0.0366 &  0.0219 \\ \midrule
CR-\textgreater{}SASRec & 0.0466 & 0.0273 & 0.0140 & 0.0078 & 0.0335 & 0.0179 \\ 
CR-\textgreater{}CR-pre & 0.0490 & 0.0298 & 0.0184 & 0.0096 & 0.0360 & 0.0211 \\
\midrule
IN-CR-OR & 0.0454 & 0.0269 &  0.0214 &  0.0117 & 0.0348 & 0.0202 \\
OR-IN-CR & 0.0464 & 0.0271 & 0.0176 & 0.0093 & 0.0348 & 0.0197 \\
OR-CR-IN & 0.0474 & 0.0269 & 0.0192 & 0.0101 & 0.0340 & 0.0185 \\
CR-IN-OR & 0.0476 & 0.0278 & 0.0220 & 0.0115 & 0.0342 & 0.0189 \\
CR-OR-IN & 0.0452 & 0.0266 & 0.0162 & 0.0085 & 0.0364 & 0.0208 \\ \bottomrule
\end{tabular}
}
\end{table}

\subsubsection{Implementation Details}

Following~\cite{wang2025msl}, we implement traditional recommenders methods by Adam \cite{kingma2014adam} optimizer with a learning rate of 0.001, an embedding dimension of 64 and a batch size of 256. 
We tune the weight decay in \{1e-4, 1e-5, 1e-6, 0\} and the dropout ratio among $[0, 0.5]$ in the step of 0.1. 
We use DROS as the backbone of LLM-enhanced methods, as it is the SOTA traditional sequential recommender.
For all LLM-enhanced and LLM-based baselines, unless otherwise specified, we use the representative LLaMA3-3B model~\cite{dubey2024llama} as the LLM backbone. We also conducted experiments on LLMs of different scales to verify the generalization of the method, and the results are shown in Table~\ref{tab:8b}. Besides, for the compared LLM-based methods, we test their methods with both language-based strategy and embedding-based strategy, and report the best ones.
We closely followed the settings suggested by their original papers and finely tuned their hyperparameters to ensure their optimum.  All methods are implemented with PyTorch and run on 8 Nvidia A800 GPUs.

For our method, all items in prompts are represented by their titles, and we adapt a recommendation projection head on the next item embedding following~\cite{xu2025slmrec}. We use the BIGRec with embedding-based prediction strategy as the backbone, because we find it performs the best at most cases. We finetune the LLM by LoRA (Low-Rank Adaptation)~\cite{hu2022lora} with a LoRA rank of 8, $\alpha$ of 16 and dropout rate of 0.05. We tune the learning rate in \{1e-3, 5e-4, 1e-4, 5e-5\}, the shallow layer number in \{1, 4, 8, 12, 16\} and the deep layer number in \{1, 2, 3, 4, 5\}. We set the max training step at 1600 and evaluate the performance per 100 steps. Early stop strategy~\cite{prechelt2002early} is used on the $NDCG@10$ with patient as 2. 


\subsection{Performance Comparison (RQ1)}
Table~\ref{tab:main} provides a comparative analysis of the performance of the proposed HatLLM method against baseline approaches. We can find that:

1) \textbf{Overall performance comparisons.} HatLLM significantly outperforms various recommenders in terms of all evaluation metrics on three real-world datasets, and surpasses the SOTA LLM-based recommender with an average improvement of 9.13\%. The results demonstrate the effectiveness of enhancing collaborative signals modeling in LLM-based recommenders. Besides, HatLLM is simple yet effective, requiring only minimal modifications to the LLM’s attention masking scheme without introducing additional overhead. Therefore, HatLLM can be seamlessly integrated as a plugin into LLM-based recommenders to enhance LLM's ability to capture collaborative signals.


2) \textbf{Compared with LLM-enhanced recommenders.}
HatLLM outperforms LLM-enhanced recommenders across all metrics and datasets. The LLM-enhanced recommendation is also a method that integrates semantic and collaborative information. However, these methods have the following limitations: (i) They do not directly leverage more powerful LLMs for recommendation, where the full potential of LLMs is not effectively utilized. 
(ii) They are difficult to train in an end-to-end manner, resulting in semantic knowledge extracted by LLMs not being well-aligned with the recommendation task.  (iii) There exists a capability and semantic gap between LLMs and traditional models, making it challenging for them to implement knowledge transfer and fusion.



3) \textbf{Compared with LLM-based recommenders.}
HatLLM exhibit substantial improvements over the LLM-based recommenders. 
Specifically, for Lite-LLM4Rec which combines two different LLMs (as introduced in Section 4.2), we observe that its performance is inferior to ours. This demonstrates the importance of the Intra-item Attention strtegy (IN) of our HatLLM, which can model the semantics within items more effectively and flexibly.  
As for LLaRA, while it shows some improvement over BIGRec, the gains remain relatively limited due to the information loss from low-dimensional collaborative embeddings and the dilution of the enhanced collaborative tokens. In contrast, our method enhances the attention mechanism to better leverage collaborative information, and the experimental results confirm that our approach achieves further performance gains over LLaRA.

Besides, Table~\ref{tab:8b} showcases the performance of the proposed HatLLM against representative LLM-based baselines on LLMs of other scales. We can find that: the proposed HatLLM outperform all representative LLM-based baselines on Llama3-1B, 3B, 8B. This result further demonstrates the effectiveness of our method and its generalization ability on LLMs of different parameter scales.

\subsection{Ablation Study (RQ2)}
We perform the following ablation study to investigate the effects of each components: 1) We test the results where shallow Intra-item Attention layers (\textbf{w/o IN}), middle orignal Attention layers (\textbf{w/o OR}) and deep cross-item Attention layers (\textbf{w/o CR}) are removed, respectively; 2) To demonstrate the effectiveness of our cross-item attention, we replace the this module with the typical traditional model SASRec (\textbf{CR->SASRec}). Specifically, we extract semantic-enhanced item embeddings from LLMs to augment traditional recommendation models. We have tested the learnable projector~\cite{su2012linear} and dimension reduction strategy~\cite{mackiewicz1993principal} for embedding transformation and report the best one; 3) We also test the performance of initial cross-item attention design (\cf eq.(4)) for comparisons (\textbf{CR->CR-pre}); 4) To demonstrate the effectiveness of our progressive attention design from fine-grained token-level to coarse-grained item-level, we permute the order of three typical layers and report their performance (\eg \textbf{"IN-CR-OR"} means using IN in the shallow layers, CR in the middle layers, and OR in the deep layers). We can observe that: 

1) Overall, removing the IN, OR, and CR modules from HatLLM leads to a significant decline in recommendation performance. 
This demonstrates the importance of HatLLM's hierarchical attention masking strategy at the shallow, middle, and deep layers, and proves the significance of simultaneously capturing semantic signals and collaborative signals in LLM-based recommendation for improving recommendation performance. 


2) Replacing the CR module with SASRec leads to a significant performance drop. This highlights the limitations of directly stacking LLMs and traditional models: (i) The mismatch in embedding spaces necessitates transformation and compression, causing information loss; (ii) The decoupled design hinders end-to-end optimization, whereas HatLLM's integrated architecture achieves superior performance through joint fine-tuning.

3) Replacing the CR module with CR-pre also leads to significant performance degradation. This demonstrates the limitations of relying solely on local attention to neighboring items: (i) Attention weights overly concentrate on tokens from adjacent items, failing to capture long-range dependencies; (ii) Since neighboring items in the sequence may have disparate semantics, excessive aggregation of local information introduces noise and harms the performance.

4) Considering changing the order of the three attention layers, from Table~\ref{tab:ab}, we can observe that the recommendation performance of HatLLM using our proposed "IN-OR-CR" order surpasses all other orders. This further demonstrates the progressive rationale behind our proposed HatLLM: using IN in the shallow layer to encode item semantics, OR in the middle layer to model contextual semantics, and CR in the deep layer to focus on capturing collaborative signals. Such the order from the fine-grained semantic token-level to coarse-grained collaborative item-level could better elicit the capabilities of LLMs.

\subsection{Hyperparameter Sensitivity (RQ3)}
\begin{figure}[t]
    \centering
    \begin{subfigure}{0.235\textwidth}
        \centering
        \includegraphics[width=\textwidth]{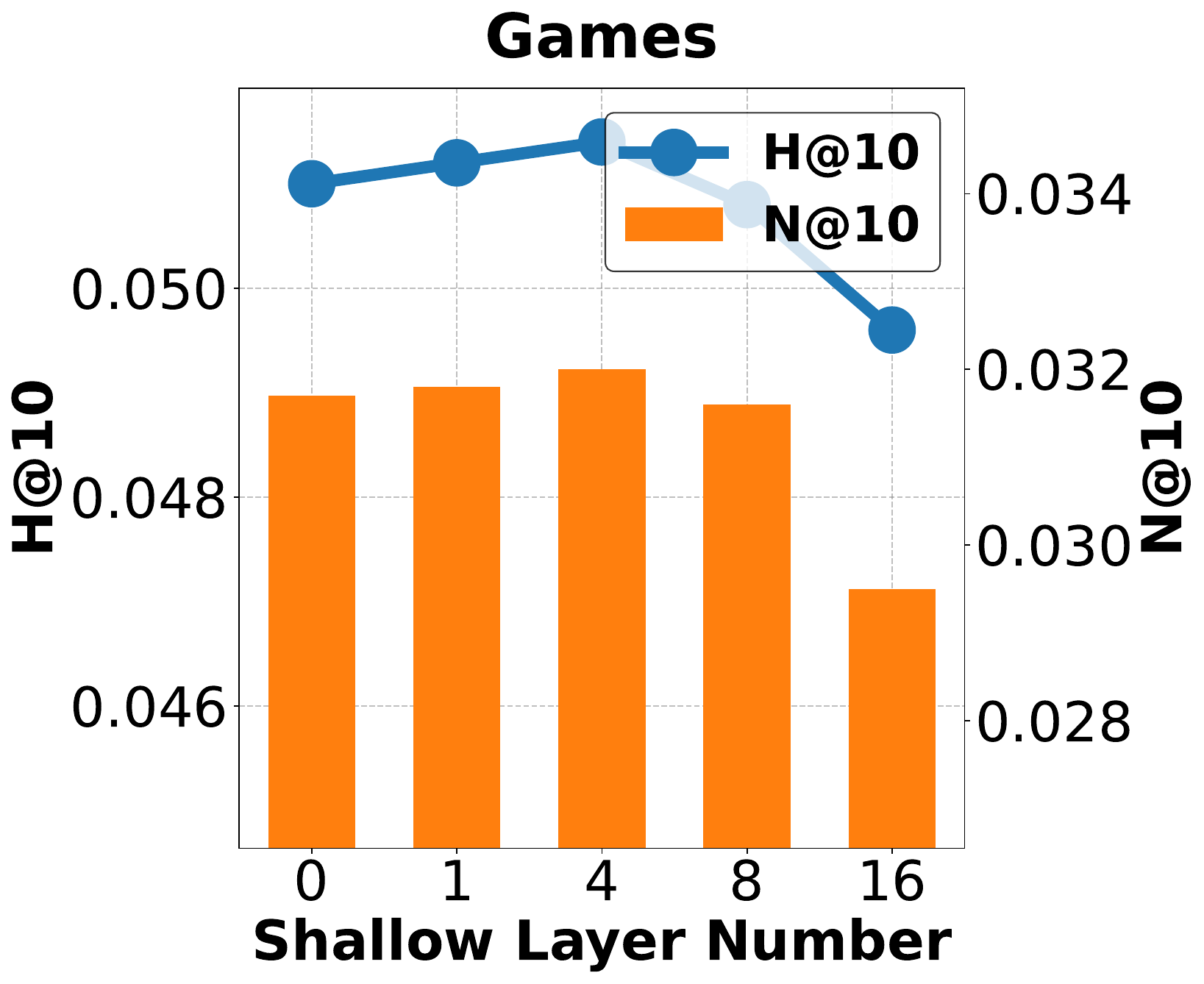}
    \end{subfigure}    
    \begin{subfigure}{0.235\textwidth}
        \centering
        \includegraphics[width=\textwidth]{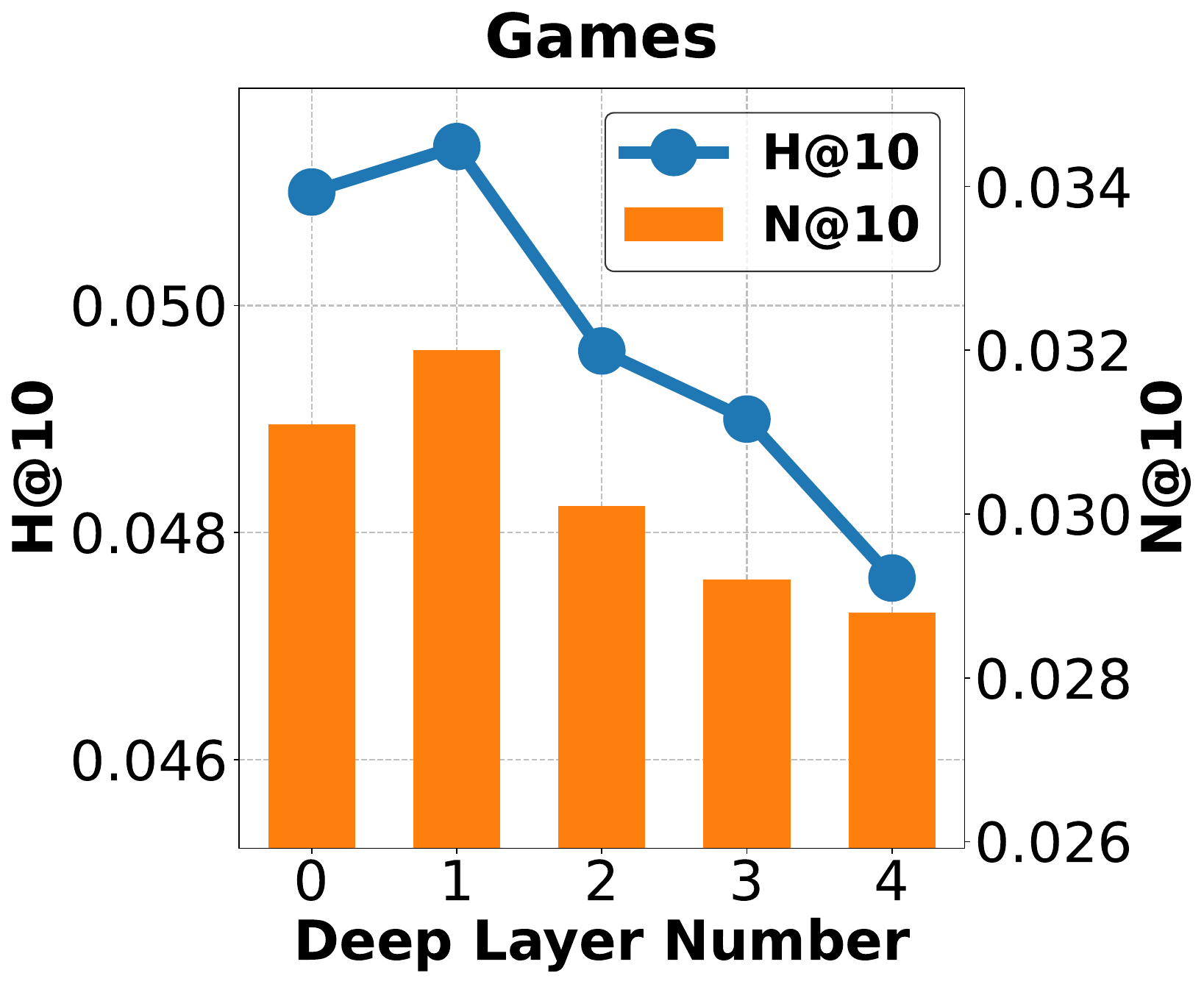}
    \end{subfigure}
    \begin{subfigure}{0.235\textwidth}
        \centering
        \includegraphics[width=\textwidth]{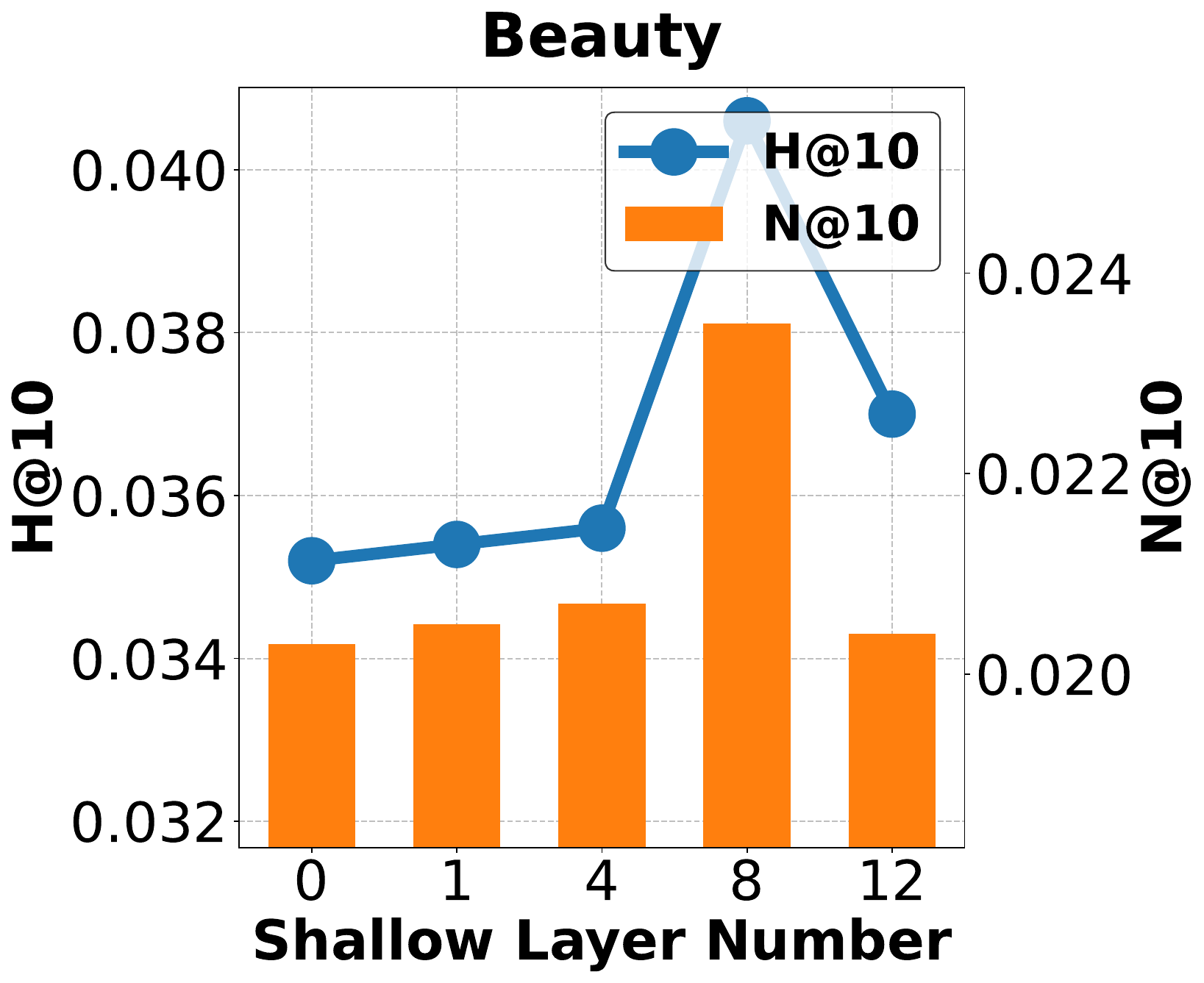}
    \end{subfigure}    
    \begin{subfigure}{0.235\textwidth}
        \centering
        \includegraphics[width=\textwidth]{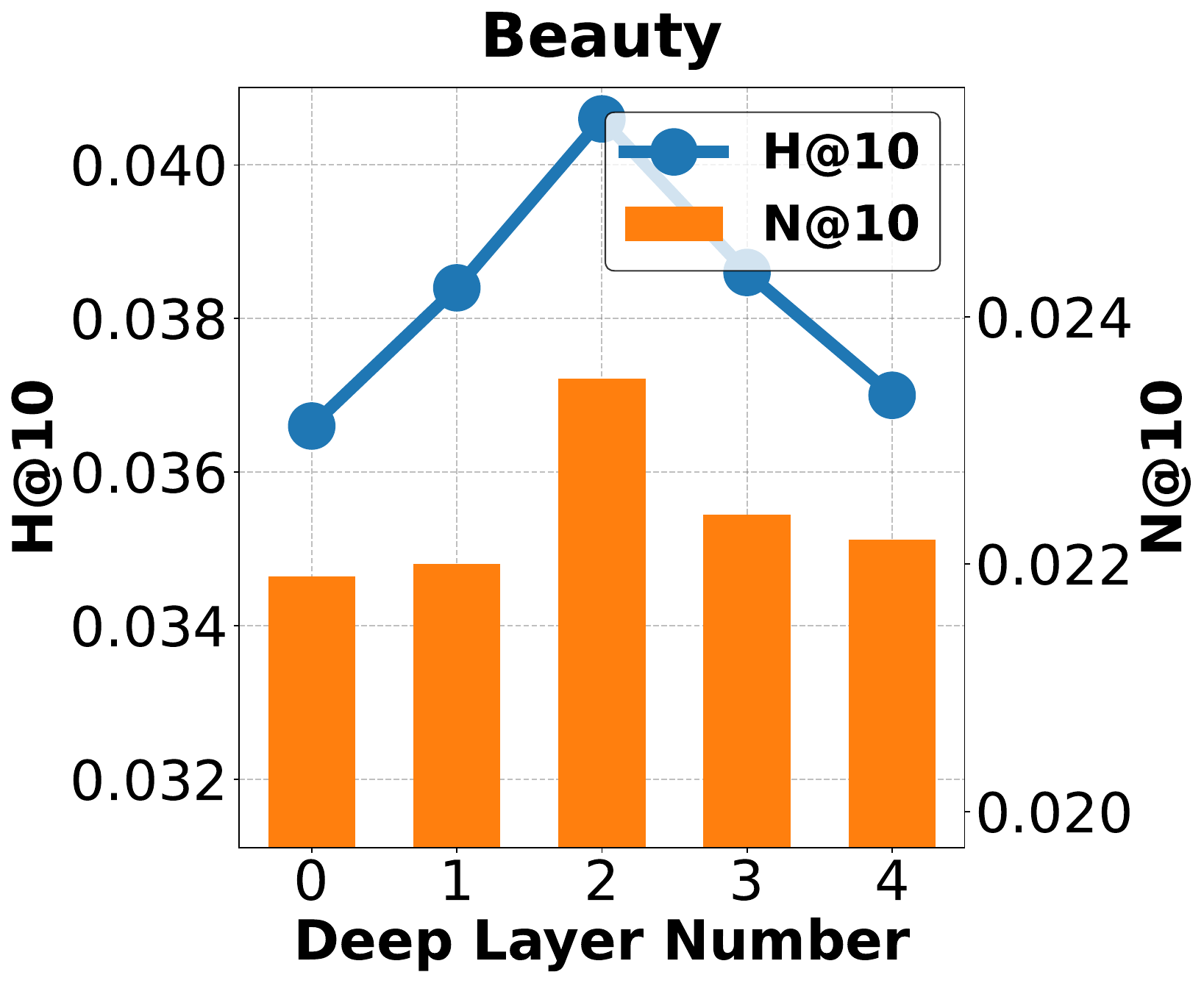}
    \end{subfigure}
    \begin{subfigure}{0.235\textwidth}
        \centering
        \includegraphics[width=\textwidth]{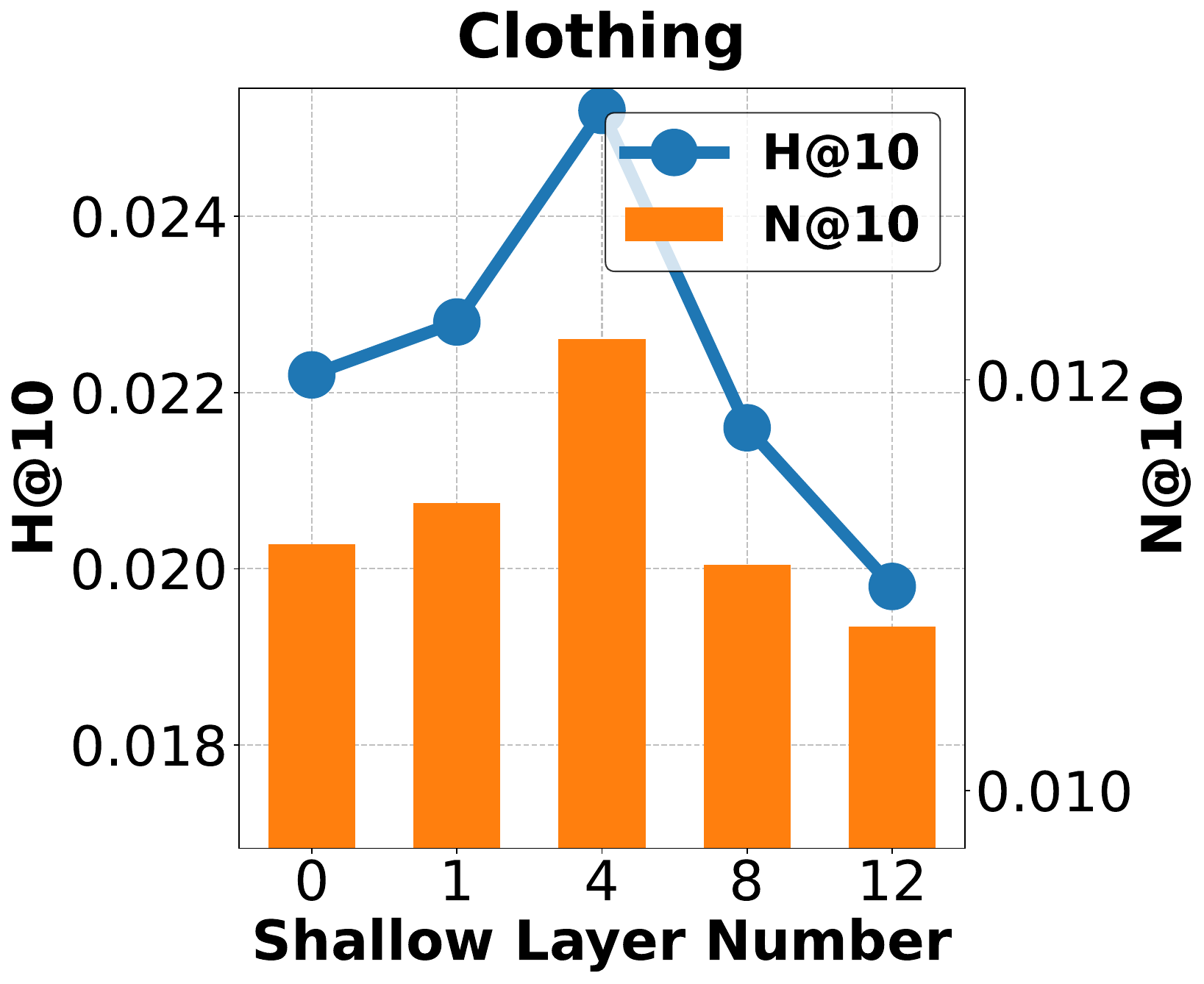}
    \end{subfigure}    
    \begin{subfigure}{0.235\textwidth}
        \centering
        \includegraphics[width=\textwidth]{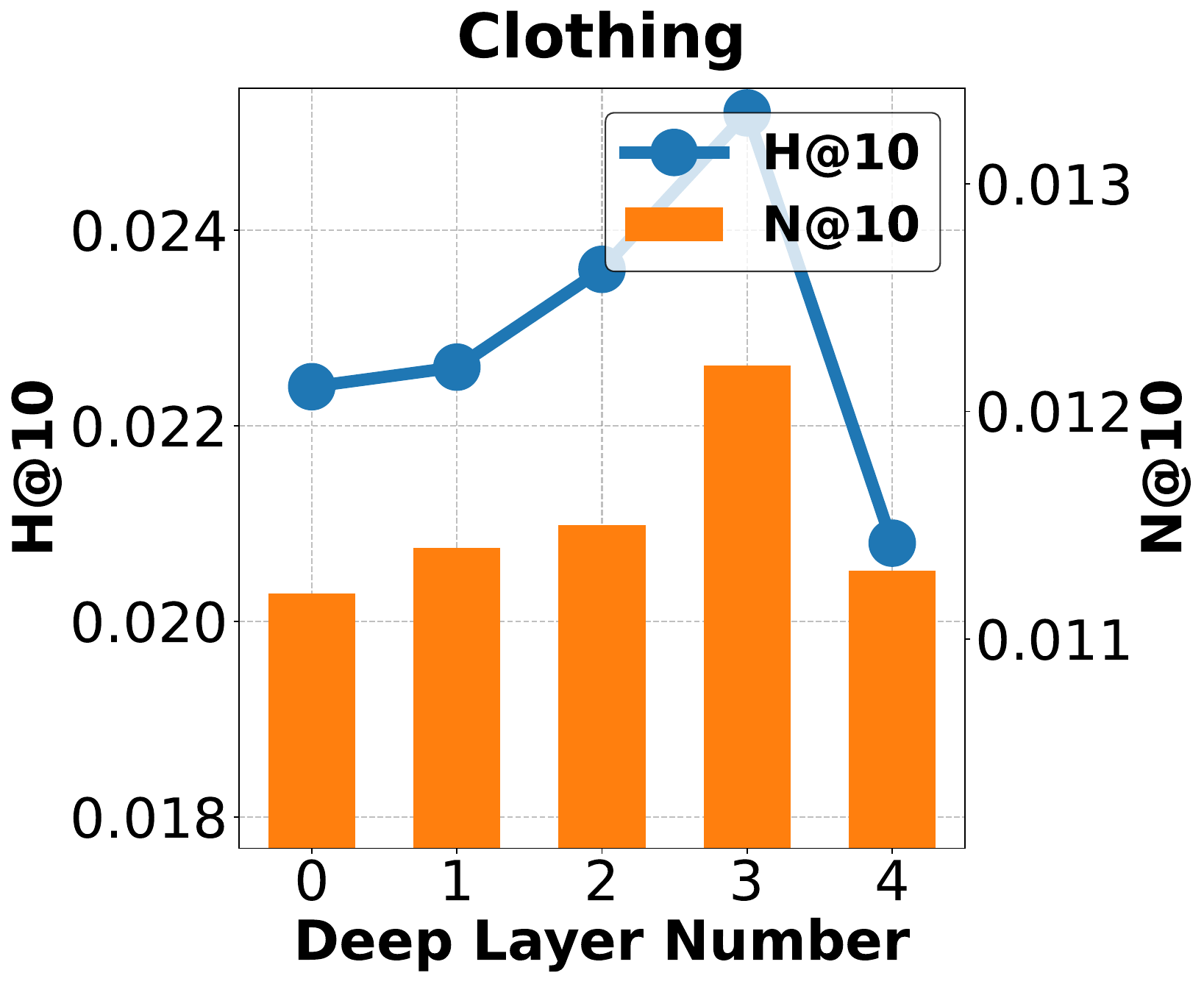}
    \end{subfigure}
    \vspace{-0.3cm}
    \caption{Hyperparameter sensitivity analysis on shallow and deep layer numbers.} 
    \label{fig:hyper}
    \vspace{-0.5cm}
\end{figure}

Figure~\ref{fig:hyper} illustrates the performance of HatLLM with different hyperparameters (\ie the shallow intra-item and deep cross-item layer numbers). We can observe that: 1) With the increase in the number of shallow layers, the model's recommendation performance first rises and then falls, with better results achieved at shallow layer numbers of 4 or 8. This indicates that the parameters of LLM used for item encoding should not be too verbose or too simple; an appropriate number of shallow LLM layers is sufficient for learning item semantics in a learnable manner. 2) The number of deep layers should not be too high, typically requiring only 1-3 layers to achieve good results. This is because deep layers in HatLLM are responsible for learning item-level collaborative information, and recent work on traditional models demonstrate stacking several transformer layers is sufficient ~\cite{kang2018self}. 

\section{Related Work}
\subsection{Sequential Recommendation}
Sequential recommendation~\cite{xie2022contrastive, chen2018sequential, chang2021sequential, li2020time} aims to predict what the users might prefer next with their historical behaviors. Existing sequential recommendation methods use sequential modeling models, such as recurrent neural networks (RNNs)~\cite{hidasi2015session}, convolutional neural networks (CNNs)~\cite{tang2018personalized} or Transformers~\cite{sun2019bert4rec,kang2018self,yang2023generic}, to model user interaction sequences. 
For example, SASRec~\cite{kang2018self} introduces attention mechanism to automatically learn the weights of different interaction items. DROS~\cite{yang2023generic} leverage distributional robust optimization (DRO) in sequential recommendation to improve the model’s robustness. 
Most current LLM for RS methods also adopt the setting of sequential recommendation~\cite{bao2023tallrec,bao2025bi,liao2024llara}.

\subsection{LLMs for Recommendation}
Large language models (LLMs) have been widely applied in RS for their powerful comprehension capabilities and extensive knowledge~\cite{li2023prompt,qin2025d2k,zhang2023recommendation,shi2024enhancing,xu2024enhancing,wang2024cela,geng2022recommendation, cui2022m6}. There are primarily two paradigms to utilize LLMs in RS:

\textbf{LLM-based recommenders.} This paradigm directly leverages pre-trained LLMs as the backbone for recommendations~\cite{zhu2023collaborative,hou2023learning,zheng2024adapting,jiang2024item,tan2024llmrecsys,chen2024softmax,liu2024once, chen2024hllm}. Early studies explored the capabilities of LLMs in zero-shot scenarios by framing the recommendation task as language prompts~\cite{gao2023chat, hou2024large, wang2023zero, liu2023first}. 
However, due to the discrepancy between the recommendation task and the pre-training language modeling task of LLMs, these methods often perform poorly.
Subsequently, the researchers explored finetuning LLMs with recommendation data to minimize the semantic gap between recommendation and natural language modeling~\cite{bao2023tallrec,lin2024bridging,rajput2023recommender,wang2025msl,kim2024large}.
An intuitive approach is \textbf{textual-based methods}, which use language modeling loss (LML) to train LLMs for predicting the next item descriptions, named . For instance, BIGRec~\cite{bao2025bi} employs instruction tuning on LLMs to generate the next item titles and proposes a grounding strategy to map them into real item titles. LLaRA~\cite{liao2024llara} injects the collaborative embeddings of traditional recommenders into LLMs to enhance textual-based recommendation. Although effective, they face challenges in inference efficiency and hallucination issues. 
To address these problems, \textbf{embedding-based methods} map the next item embedding predicted by the LLM to the real item space through a projection head to directly generate recommendation results. For example, E4SRec~\cite{li2023e4srec} and SLMRec~\cite{xu2025slmrec} organize the item embeddings from traditional recommenders into prompts to predict the next item embedding, and trains them through a learnable recommendation projection head with recommendation loss. 

Although effective, existing LLM-based recommenders all employ token-level processing for user behavior modeling. Our experiments observe that token-level processing will weaken the ability of LLMs to capture significant collaborative signals, which is critical for RS. Our proposed HatLLM effectively addresses this issue through hierarchical attention masks, further enhancing the reasoning capabilities of LLMs.

\textbf{LLM-enhanced recommenders.} This paradigm primarily leverages the rich knowledge and reasoning capabilities of LLMs to enhance traditional recommender models~\cite{ren2024enhancing,ren2024representation,sun2024large,wang2024can,ren2023representation,wei2024llmrec,liu2024llm}. For instance, KAR~\cite{xi2024towards} utilizes LLMs to enrich user and item information, encoding them as semantic vectors to augment traditional models. LLM-CF~\cite{sun2024large} employs LLMs to generate  Chain-of-Thought (CoT) for recommendation data to enhance traditional models. DLLM2Rec~\cite{cui2024distillation} uses knowledge distillation to extract student-friendly knowledge feom LLMs for traditional recommenders enhancement. The main challenge of this paradigm lies in the significant capability and mechanism gap between LLMs and traditional recommender models, which hinders effective knowledge transfer. This is not the focus of our research and readers may refer to the surveys~\cite{liu2024large,zhao2024recommender} for more details.

\section{Conclusion}
In this paper, we reveal that the widely-used token-level processing of LLM-based recommenders will inherently limits LLMs’ ability to capture the significant collaborative signals within cross-item correlations. To overcome this challenge, we propose a novel hierarchical attention masking strategy, HatLLM, specifically tailored for this challenge. HatLLM contains intra-item attention (IN) in shallow layers, original attention (OR) in middle layers and cross-item attention (CR) in deep layers, aiming to empowers LLMs to jointly capture token-level and item-level dependencies for improved recommendation. Extensive experiments on three real-world datasets demonstrate the effectiveness of our method. In the future, it would be interesting to explore more advanced attention mechanism that could capture both semantic and collaborative signals on common layers. 

\bibliographystyle{ACM-Reference-Format}
\bibliography{sample-base}

\clearpage
\appendix
\section{Appendix}


\subsection{Baselines}
This section provides a brief introduction to each baseline model used in our experiments.
\begin{itemize}[leftmargin=*]
\item \textbf{SASRec} (ICDM'18)~\cite{kang2018self} is a transformer-based recommender widely used in sequential recommendation. It leverages self-attention to capture collaborative signals in user interaction sequences and uses cross-entropy loss as the optimization objective.
\item \textbf{DROS} (SIGIR'23)~\cite{yang2023generic} proposes a Distributionally Robust Optimization framework (DRO) for sequential recommendation to address distribution shifts between training and testing data caused by dynamic environments. It is model-agnostic, applicable to various backbone recommenders.
\item \textbf{KAR} (RecSys'24)~\cite{xi2024towards} augments recommender systems with LLM-derived knowledge (user preferences and item facts) via factorization prompting and hybrid-expert adaptation, boosting performance while enabling efficient deployment.
\item \textbf{LLM-CF} (CIKM'24)~\cite{sun2024large} enhances collaborative by distilling LLMs' world knowledge and reasoning capabilities through in-context learning. It introduces an efficient instruction-tuning method to improve recommendation performance while preserving LLMs' general abilities.
\item \textbf{DLLM2Rec} (RecSys'24)~\cite{cui2024distillation} proposes a knowledge distillation framework that transfers capabilities from large language model to traditional recommenders via importance-aware ranking distillation and collaborative embedding distillation.
\item \textbf{E4SRec} (arXiv'23)~\cite{li2023e4srec} proposes an ID-compatible LLM framework for sequential recommendation, which directly processes item ID sequences while ensuring generated outputs stay within valid candidate sets. 
\item \textbf{Lite-LLM4Rec} (arXiv'24)~\cite{wang2024rethinking} is a lightweight LLM-based sequential recommender that replaces beam search with an efficient item projection head and adopts a hierarchical LLM structure to reduce computational overhead.
\item \textbf{SLMRec} (ICLR'25)~\cite{xu2025slmrec} proposes a lightweight sequential recommendation framework that distills knowledge from LLMs into small language models. It theoretically analyzes why small language models suffice for sequential recommendation and demonstrates compatibility with quantization/pruning techniques.
\item \textbf{BIGRec} (TORS'25)~\cite{bao2025bi} is a two-step grounding framework that first adapts LLMs to the recommendation space by fine-tuning them to generate meaningful item tokens, then maps these tokens to actual items.
\item \textbf{LLaRA} (SIGIR'24)~\cite{liao2024llara} bridges conventional recommenders and LLMs by aligning ID embeddings with textual features via hybrid prompting, which can inject collaborative signals into LLMs and improve the recommendation performance.
\end{itemize}

\end{document}